\documentclass[prl,twocolumn,showpacs]{revtex4}
\usepackage{times}
\usepackage{amsmath, amssymb}

\def\kbar{\protect\@kbar}
\def\@kbar{%
\relax \bgroup
\def\@tempa{\hbox{\raise.73\ht0
\hbox to0pt{\kern.25\wd0\vrule width.5\wd0
height.1pt depth.1pt\hss}\box0}}%
\mathchoice{\setbox0\hbox{$\displaystyle k$}\@tempa}%
{\setbox0\hbox{$\textstyle k$}\@tempa}%
{\setbox0\hbox{$\scriptstyle k$}\@tempa}%
{\setbox0\hbox{$\scriptscriptstyle k$}\@tempa}%
\egroup}

\input epsf

\begin{document}

\title{Experimental observation of high-order quantum accelerator modes}
\author{S. Schlunk,$^{1}$ M.B. d'Arcy$,^{1}$ S.A.
Gardiner,$^{1}$ and G.S. Summy$^{1,2}$}

\affiliation{$^{1}$Clarendon Laboratory, Department of Physics,
University of Oxford, Parks Road, Oxford, OX1 3PU, United Kingdom
\\ $^2$Department of Physics, Oklahoma State University,
Stillwater, Oklahoma, 74078-3072}
\date{\today}

\begin{abstract}
Using a system consisting of a freely falling cloud of cold cesium
atoms periodically kicked by pulses from a vertical standing wave
of laser light, we present the first experimental observation of
high-order quantum accelerator modes. This confirms the recent
prediction by Fishman, Guarneri, and Rebuzzini [Phys.\ Rev.\
Lett.\ {\bf 89}, 084101 (2002)]. We also show how these
accelerator modes can be identified with the stable regions of
phase space in a classical-like chaotic system, despite their
intrinsically quantum origin.
\end{abstract}


\maketitle

The search for signatures of chaos and stability in quantum
systems whose classical analogs can exhibit chaotic dynamics is an
area of intense current theoretical and experimental interest. The
motivation for such investigations is twofold. Firstly, the study
of the way in which complex classical behaviour has its origin in
the quantum regime helps in understanding the operation of the
quantum-classical correspondence principle and thus the physical
processes that are crucial in determining observed macroscopic
behaviour, particularly when this behaviour is unpredictable and
chaotic. Secondly, the quantum dynamics of such systems are of
considerable interest in their own right, especially when the
systems behave in a peculiarly non-classical manner.

The study of quantum chaos and stability has focused on both
energetic and dynamical properties of systems
\cite{casati2000,gutzwiller90,haake2001,reichl92}. Approaches to
the classification of quantum systems' behaviour have ranged from
the highly mathematical (e.g. trace formulae \cite{gutzwiller90})
through the statistical (energy spectra \cite{haake2001}) to the
more phenomenological (energy and momentum transfer to ensembles
of particles \cite{reichl92}). It is the latter approach which is
most appealing as a philosophy to guide experimental
investigations, and has underpinned the work presented in this
Letter. Resonant, stable behaviour in quantum systems often
depends on precise fulfilment of matching conditions between
periodic forcing of a system and its own natural frequency. This
is in stark contrast to the looser matching which is generally
required in the corresponding classically chaotic system
\cite{fishman2002}. The quantum resonances \cite{izrailev79}
observed in the $\delta$-kicked rotor \cite{lichtenberg92,moore95}
represent an excellent example of this dichotomy between the
quantum and the classical. These resonances are characterised by
the steady transfer of momentum to the system, which in the
atom-optical case \cite{oskay2000,darcy2001b} manifests itself as
a symmetric broadening of the atomic momentum distribution for
special values of the driving frequency of the potential. This
phenomenon is therefore interesting in terms of both the
motivations outlined above.

In this Letter, we report the results of experiments using an
atom-optical realization \cite{klappauf98} of the $\delta$-kicked
accelerator \cite{oberthaler99,godun2000,schlunk2002,darcy2001a}.
This is realized when pulses from a vertical standing wave of
laser light are applied to freely falling laser-cooled atoms. The
Hamiltonian for this system is given by
\begin{equation}
\hat{H} = \frac{\hat{p}^{2}}{2m} + mg\hat{z} - \hbar \phi_{d} [1 +
\cos(G\hat{z})]\sum_{n}\delta(t-nT), \label{eq:hamiltonian}
\end{equation}
which is related to the classically chaotic $\delta$-kicked rotor
by the addition of a static linear potential. Here $\hat{z}$ is
the position, $\hat{p}$ the momentum, $m$ the particle mass, $g$
the gravitational acceleration, $t$ the time, $T$ the pulse
period, $G =2\pi/\lambda_{\mbox{\scriptsize spat}}$, where
$\lambda_{\mbox{\scriptsize spat}}$ is the spatial period of the
potential applied to the atoms, and $\hbar \phi_{d}$ quantifies
the depth of this potential. The classical dynamics of this system
are qualitatively similar to those of the $\delta$-kicked rotor,
whereas the quantum dynamics are quite distinct. For example, the
quantum accelerator modes recently discovered in this system
\cite{oberthaler99} are not observed in the corresponding
classical system. The modes are characterised by the asymmetric
transfer of a fixed momentum impulse per kick to approximately
$20$\% of the initial ensemble of laser-cooled atoms. In a recent
analysis of this phenomenon by Fishman, Guarneri, and Rebuzzini
(FGR) \cite{fishman2002}, quantum accelerator modes were
interpreted as being a resonant type of behavior, closely related
to the quantum resonances in the $\delta$-kicked rotor. This
analysis also led to the fascinating prediction of the existence
of whole families of higher-order quantum accelerator modes. These
were shown to correspond to higher-order fixed points centered on
systems of islands in a (pseudo)classical phase space. In this
Letter we report the first experimental observation of these
higher-order accelerator modes, finding excellent quantitative
agreement with the analysis of Ref.\ \cite{fishman2002}.

The kicking potential acts on the atoms as a phase grating that
induces a phase modulation of amplitude $\phi_{d}$ to their de
Broglie waves. Hence the effect of a pulse on a plane wave is to
cause diffraction into a series of momentum states separated by
the grating recoil $\hbar G$. Between consecutive pulses these
states accumulate a phase related to their kinetic energy. This
phase evolution is determined by the value of $T$, which will
therefore govern the type of dynamics exhibited by the system. As
in Refs.\ \cite{darcy2001a,schlunk2002}, we use scaled position
and momentum variables $\chi = Gz$ and $\rho = GTp/m$. An
effective scaled, dimensionless Planck constant $\kbar = \hbar
G^{2} T/m$ is then defined through the relation $\kbar =
-i[\hat{\chi},\hat{\rho}]$. This parameter, together with
$\phi_{d}$ and $\gamma=gGT^{2}$ (which accounts for the effect of
gravity) fully describe the quantum dynamics of the
$\delta$-kicked accelerator. When quantum resonances occur in the
$\delta$-kicked rotor ($\gamma = 0$), the phase difference
accumulated between momentum states separated by $\hbar G$ from
one pulse to the next is equal to an integer multiple of $2\pi$.
For a state of zero initial momentum, this is the case for values
of $T$ corresponding to $\kbar = 4\pi \ell$ where $\ell  \in
\mathbb{Z}$. This rephasing is analogous to the Talbot effect in
optics \cite{berry99}, and so we speak of these resonances as
occurring at integer multiples of the Talbot time $T_{{\mathrm T}}
= 4\pi m/\hbar G^{2}$ \cite{godun2000}. For a continuous initial
distribution of momenta, such as in a cold atomic ensemble,
resonant behavior is observed for $\kbar = 2\pi \ell$, i.e.\ at
integer multiples of the half-Talbot time, $T_{1/2}$
\cite{darcy2001a,darcy2001b,oskay2000}. Close to these values of
$T$, quantum accelerator modes are found in the $\delta$-kicked
accelerator \cite{darcy2001a}.

In our experimental realization of the quantum $\delta$-kicked
accelerator, about $10^{7}$ caesium atoms are trapped and cooled
in a magneto-optic trap (MOT) to a temperature of $5 \mu$K,
yielding a Gaussian momentum distribution with FWHM $6\hbar G$.
The atoms are then released from optical molasses and, falling
freely under gravity, are exposed to pulses from a vertical
standing wave of off-resonant laser light that is $20$\thinspace
GHz red-detuned from the $6^{2}\mbox{S}_{1/2} \rightarrow
6^{2}\mbox{P}_{1/2}$, $(F=4 \rightarrow F'=3)$ D1 transition.
Hence $\lambda_{\mbox{\scriptsize spat}}=447$\thinspace nm and
$T_{1/2}=66.7\mu$s. The intensity of the light in each pulse is
approximately $1\times 10^{8}$ W/cm$^{2}$, and the duration of
each pulse is $500$\thinspace ns. Through the action of the ac
Stark shift, these pulses result in $\delta$-function-like
applications of a spatially periodic potential to the atoms, with
$\phi_{d} = \Omega^{2}t_{p}/8\delta_{L}$. Here $\Omega$ is the
Rabi frequency, $t_{p}$ the duration of each (square) pulse and
$\delta_{L}$ the detuning from the D1 transition. Both the trapped
atom density distribution and the standing light wave intensity
profile are Gaussian, with full width at half maximum (FWHM) of
$1$\thinspace mm. The resulting mean value of $\phi_{d}$ is around
$0.8\pi$. After application of the diffracting pulses, the atoms
fall through a sheet of light resonant with the
$6^{2}\mbox{S}_{1/2}\rightarrow 6^{2}\mbox{P}_{3/2}$, $(F=4
\rightarrow F''=5)$ D2 transition, located $0.5$m below the point
of release, and their momentum distribution is measured by a
time-of-flight technique with a resolution of $\sim\hbar G$. For
more details regarding our experimental setup see Refs.\
\cite{godun2000,darcy2001a,schlunk2002}.

The approach used by FGR \cite{fishman2002} accounts for the
observed acceleration of atoms participating in a quantum
accelerator mode in terms of stable fixed points in a map for a
classical point particle. The validity of this map can be
justified asymptotically by the closeness of $\kbar$ to integer
multiples of $2\pi$. The pseudoclassical limit can be described by
$\epsilon  = (\kbar-2\pi \ell) \rightarrow 0$ (hence the
description `pseudoclassical'), or equivalently by $(\ell -
T/T_{1/2})\rightarrow 0$. In an appropriately transformed frame
\cite{fishman2002}, the map is given by \cite{schlunk2002}
\begin{eqnarray}
\tilde{\rho}_{n+1} &=& \tilde{\rho}_{n} - \tilde{k}\sin(\chi_{n})
-\mbox{sign}(\epsilon)\gamma, \label{eq:rhomap}\\ \chi_{n+1} &=&
\chi_{n} + \mbox{sign}(\epsilon)\tilde{\rho}_{n+1},
\label{eq:chimap}
\end{eqnarray}
where $\tilde{\rho}= \rho \epsilon/\kbar$, and
$\tilde{k}=\phi_{d}|\epsilon|$. With the correct initial
conditions, iteration of Eqs.\ (\ref{eq:rhomap}) and
(\ref{eq:chimap}) yields systems of accelerator orbits. These are
stable fixed points centered on islands in the pseudoclassical
phase space. To yield an observable quantum accelerator mode an
island system must be sufficiently large, in terms of total
phase-space area. Thus, in a classical sense, the islands must
encompass sufficient phase space density for the accelerator mode
to be measurable. Furthermore, the islands must be large, or at
least comparable to $|\epsilon|$ (which takes the place of $\hbar$
as a measure of the size of a minimal `quantum phase space' cell)
for a point particle-like description of the quantum accelerator
mode dynamics to be appropriate. We show in this Letter that when
these requirements are satisfied, the momentum gain predicted by
the analysis of FGR agrees very well with experiment, even when
$\kbar$ is not extremely close to a resonant value. This can be
understood as being due to the relevant dynamics taking place in
stable regions of the pseudoclassical phase space, where
semiclassical analyses can generally be expected to work
reasonably well \cite{zurek94}.

FGR \cite{fishman2002} classify accelerator orbits (and thus
quantum accelerator modes) by the order $\mathbf{p}$ of the fixed
point (i.e.\ how many pulse periods it takes before cycling back
to the initial point in the reduced phase space cell) and the
jumping index $\mathbf{j}$ (related to how many units of the
momentum period of phase space are imparted to the accelerating
atoms per cycle). Particles in a pseudoclassical
$(\mathbf{p},\mathbf{j})$ mode with an initial momentum of
$q_0\hbar G$ have, after $N$ kicks, a momentum (in units of $\hbar
G$) given by
\begin{equation}
  q\simeq q_0  + \frac{N}{|\ell - T/T_{1/2}|}\left[
  \frac{\mathbf{j}}{\mathbf{p}}
  + \mbox{sign}(\ell - T/T_{1/2})\frac{\gamma}{2\pi}\right],
  \label{eq:pj modes}
\end{equation}
in a frame accelerating with gravity. Only atoms of certain
initial momenta will be accelerated \cite{godun2000}; ideal values
of $q_{0}$ and initial position can be determined analytically
\cite{fishman2002}. As our molasses-cooled atomic ensemble extends
over many phase space cells, all such conditions can be satisfied.
All atoms fulfilling these conditions will receive the same
momentum transfer. Since the mean initial momentum is 0, the
central momentum of the observed accelerator mode is well
described by Eq.\ (\ref{eq:pj modes}) with $q_{0}=0$.

\begin{figure}[tbp]
\begin{center}
\mbox{ \epsfxsize 3.3in\epsfbox{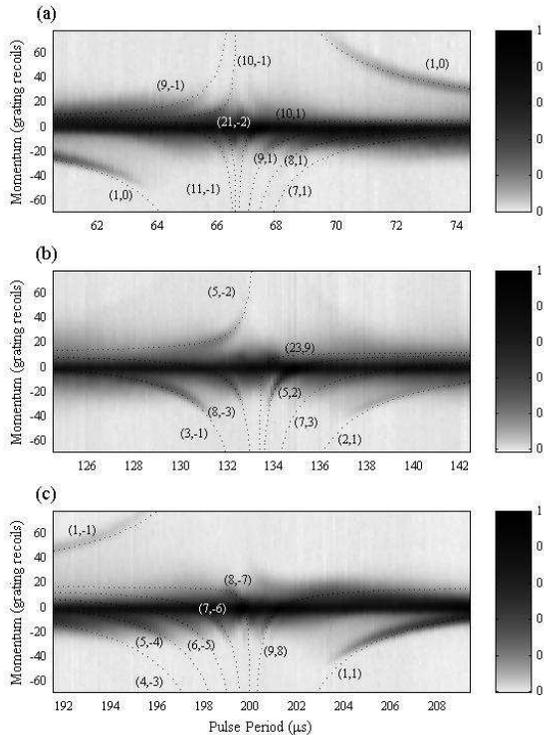}}
\end{center}
\caption{Pseudocolor plots of the variation with $T$ of the
experimental momentum distribution after 30 kicks, in a frame
falling freely with gravity. The value of  $T$ was varied around
(a) $T_{1/2}$, (b) $T_{\mathrm T}=2T_{1/2}$, and (c) $3T_{1/2}$,
in steps of $0.128\mu$s. The overlaid dotted lines indicate the
predicted momenta [Eq.\ (\ref{eq:pj modes})] of selected quantum
accelerator modes, labelled by $({\mathbf{p}},{\mathbf{j}})$.}
\label{Fig:30kicks}
\end{figure}

To search for high-order accelerator modes, we measured the
momentum distribution after a fixed number of pulses for a range
of $T$ in the vicinity of the first three integer multiples of
$T_{1/2}$. Figure \ref{Fig:30kicks} displays the experimental
momentum distributions after 30 pulses for values of $T$ in the
region of (a) $T_{1/2}$ ($T$ = $60.5\mu$s to $74.5\mu$s), (b)
$T_{\mathrm{T}}=2T_{1/2}$ ($124.5\mu$s to $142.5\mu$s), and (c)
$3T_{1/2}$ ($191.5\mu$s to $209.5\mu$s). The dotted curves
indicate the theoretical predictions of Eq.\ (\ref{eq:pj modes}).
There is some disagreement for very large momenta, particularly
large negative momenta. At these momenta, the atoms have left the
Raman-Nath regime \cite{darcy2001a}, and move so quickly that they
travel a significant fraction of the standing wave period
$\lambda_{\mbox{\scriptsize spat}}$ during a pulse. These atoms
experience a spatially averaged potential, so it is no longer
appropriate to speak of $\delta$-function-like kicks. Such an
effect will be stronger for atoms accelerated in the negative
direction ({\em with}\/ gravity) than in the positive ({\em
against}\/ gravity).

Certain of the more slowly accelerating (higher-order) quantum
accelerator modes can only be resolved after applying a larger
number of pulses than used in  Fig.\ \ref{Fig:30kicks}. To observe
the emergence of several such modes, the value of $T$ was scanned,
for larger pulse numbers, in the region of $T_{\mathrm{T}}$.
Figure \ref{Fig:T Talbot} shows the experimental momentum
distributions after (a) 60, (b) 90, (c) 120, and (d) 150 pulses.
Overlaid white lines indicate the predictions of Eq.\ (\ref{eq:pj
modes}). In Fig.\ \ref{Fig:T Talbot}(a) we can now identify the
(13,-5) and (23,9) modes. After 90 kicks [Fig.\ \ref{Fig:T
Talbot}(b)] the (18,-7) accelerator mode is emerging, whereas the
momenta of the (2,1) and the (5,-2) modes have grown beyond the
measurable range. In Figs.\ \ref{Fig:T Talbot}(c) and \ref{Fig:T
Talbot}(d) the atoms have received yet more momentum, and the
(3,-1) mode is no longer visible. Note also that some of the
quantum accelerator modes seem to `fade' and become diffuse with
time; this effect is not predicted by the pseudoclassical model
and may be due to tunneling \cite{fishman2002}.

\begin{figure}[tbp]
\begin{center}
\mbox{ \epsfxsize 3.3in\epsfbox{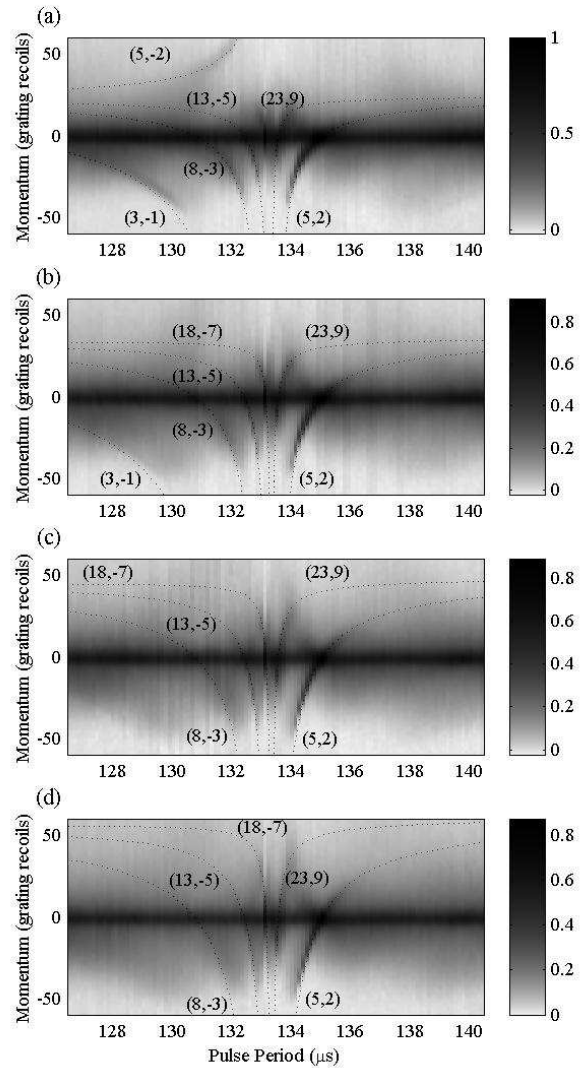}}
\end{center}
\caption{Experimental momentum distributions for different pulse
numbers as $T$ is varied in the vicinity of the Talbot time $T =
T_{\mathrm T}$, from $124.5 \mu$s to $142.5 \mu$s in steps of
$0.128 \mu$s. The total pulse number is (a) 60, (b) 90, (c) 120,
and (d) 150. The overlaid dotted lines indicate the predicted
momenta [Eq.\ (\ref{eq:pj modes})] of selected quantum accelerator
modes, labelled by $({\mathbf{p}},{\mathbf{j}})$.} \label{Fig:T
Talbot}
\end{figure}

\begin{figure}[tbp]
\begin{flushleft}
\mbox{ \epsfxsize 3.3in\epsfbox{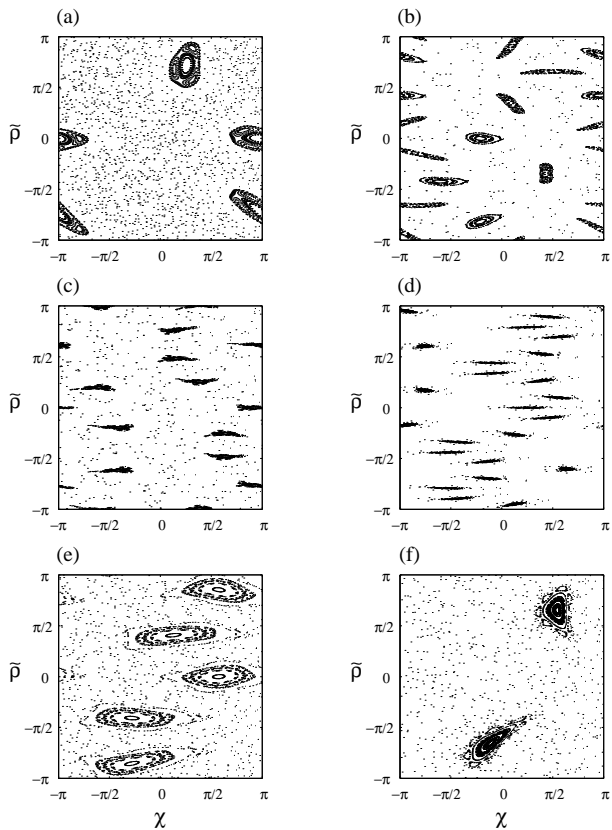}}
\end{flushleft}
\caption{Phase space plots produced by the (pseudo)classical map
of Eqs.\ (\ref{eq:rhomap}) and (\ref{eq:chimap}) for values of $T$
close to the Talbot time $T_{\mathrm{T}}$. The islands correspond
to the following quantum accelerator modes: (a) $T=130.0\mu$s, and
$(\mathbf{p},\mathbf{j}) = (3,-1)$; (b) $T=132.2\mu$s,
$(\mathbf{p},\mathbf{j}) = (5,-2)$ (shorter, rounder islands) and
(8,-3) (thin, elongated islands); (c) $T=132.8\mu$s,
$(\mathbf{p},\mathbf{j}) = (13,-5)$; (d) $T=133.5\mu$s,
$(\mathbf{p},\mathbf{j}) = (23,9)$; (e) $T=134.2\mu$s,
$(\mathbf{p},\mathbf{j}) = (5,2)$; and (f) $T=139.4\mu$s,
$(\mathbf{p},\mathbf{j}) = (2,1)$. We have clustered the initial
conditions around the fixed points corresponding to accelerator
orbits to highlight the structure of the island system of
interest. \label{Fig:phasespace}}
\end{figure}

We now explicitly connect the experimentally observed higher-order
quantum accelerator modes around $T_{\mbox{\scriptsize T}}$, as
displayed in Figs.\ \ref{Fig:30kicks}(b) and \ref{Fig:T Talbot},
with their corresponding island systems in pseudoclassical phase
space. Figure \ref{Fig:phasespace} shows stroboscopic phase space
plots, generated numerically by repeated iterations of Eqs.\
(\ref{eq:rhomap}) and (\ref{eq:chimap}), for different values of
$T$ around $T_{\text{T}}$. The island systems can be identified
with the experimentally observed quantum accelerator modes.
Comparing Figs.\ \ref{Fig:T Talbot} and \ref{Fig:phasespace}, we
see that the appearance and disappearance of the quantum
accelerator modes and of the stable island systems as $T$ is
varied coincide. Figure \ref{Fig:phasespace}(a) ($T=130.0\mu$s)
shows the three large islands corresponding to the (3,-1) quantum
accelerator mode. In Fig.\ \ref{Fig:phasespace}(b) it is possible
to observe the coexistence of a (5,-2) and a (8,-3) island system
at $T=132.2\mu$s, which can be seen to be consistent with the
experimental results in Figs.\ \ref{Fig:30kicks}(b) and \ref{Fig:T
Talbot}(a).  Interestingly, these yield  simultaneous momentum
transfer in opposite directions, promising application as a
beam-splitting technique. Figure \ref{Fig:phasespace}(c)
($T=132.8\mu$s) shows the emergence of a (13,-5) accelerator
orbit, while in Fig.\ \ref{Fig:phasespace}(d) we see a complex
(23,9) island system at $T=133.5\mu$s (just greater than
$T_{\mbox{\scriptsize T}}$). This nevertheless appears to
correspond to a fairly robust quantum accelerator mode that is
clearly visible in each of the subplots of Fig.\ \ref{Fig:T
Talbot}. Moving further away from $T_{\mbox{\scriptsize T}}$, in
Figs.\ \ref{Fig:phasespace}(e) and \ref{Fig:phasespace}(f)
($T=134.2\mu$s and $139.4\mu$s, respectively), we again observe
comparatively simple orbits, (5,2) and (2,1), respectively.

In conclusion, we have successfully observed a multitude of
quantum accelerator modes of up to 23rd order, and connected them
to the periodic orbits of a classical map. This was derived by FGR
\cite{fishman2002} as a pseudoclassical limit of the underlying
quantum dynamics when the pulse period approaches certain
resonance times. Linking this theory with our experiment, we have
successfully performed quantum accelerator mode spectroscopy.
Confirmation of the validity of such a theoretical approach
promises new avenues for investigation of quantum-classical
correspondence in a chaotic context. Furthermore, the  efficient
momentum transfer occurring in these atomic dynamics is of great
intrinsic interest. We have recently demonstrated quantum
accelerator modes to be formed coherently \cite{schlunk2002}, and
the simultaneous existence of quantum accelerator modes in
opposite momentum directions could be applied as a beam-splitter
for large-area atom interferometry \cite{berman97}.

We thank R. Bach, K. Burnett, S. Fishman, I. Guarneri, L.
Rebuzzini, and S. Wimberger for stimulating discussions. We
acknowledge support from the UK EPSRC, the Paul Instrument Fund of
The Royal Society, the EU as part of the TMR `Cold Quantum Gases'
network, contract no.\ HPRN-CT-2000-00125, the DAAD (S.S.), and
the Royal Commission for the Exhibition of 1851 (M.B.d'A.).

\end{document}